\documentclass[aps,prl,twocolumn,showpacs,superscriptaddress,groupedaddress]{revtex4}  

\usepackage{graphicx}  
\usepackage{dcolumn}   
\usepackage{bm}      
\usepackage{amssymb}  
\usepackage{color}

\usepackage{epstopdf}

\begin{document}

\title{The Quantum Hall Effect with Wilczek's charged magnetic flux tubes instead of electrons}
\date{\today}

\author{Marija Todori\'{c}}
\affiliation{Department of Physics, Faculty of Science, 
	University of Zagreb, Bijeni\v{c}ka c. 32, 10000 Zagreb, Croatia}

\author{Dario Juki\'c}
\affiliation{Faculty of Civil Engineering, University of Zagreb, A. Ka\v{c}i\'{c}a  Mio\v{s}i\'{c}a 26, 10000 Zagreb, Croatia}

\author{Danko Radi\'c}
\affiliation{Department of Physics, Faculty of Science, 
University of Zagreb, Bijeni\v{c}ka c. 32, 10000 Zagreb, Croatia}

\author{Marin Solja\v{c}i\'c}
\affiliation{Department of Physics, Massachusetts Institute of Technology,
Cambridge, Massachusetts 02139, USA}

\author{Hrvoje Buljan}
\affiliation{Department of Physics, Faculty of Science, 
University of Zagreb, Bijeni\v{c}ka c. 32, 10000 Zagreb, Croatia}

\begin{abstract}
Composites formed from charged particles and magnetic flux tubes, proposed by Wilczek, are one model for anyons - particles obeying fractional statistics. 
Here we propose a scheme for realizing charged flux tubes, in which a charged object with an intrinsic magnetic dipole moment is placed between two semi-infinite blocks of a high permeability ($\mu_r$) material, and the images of the magnetic moment create an effective flux tube. 
We show that the scheme can lead to a realization of Wilczek's anyons, when a two-dimensional electron system, which exhibits the integer quantum Hall effect (IQHE), is sandwiched between two blocks of the high-$\mu_r$ material with a temporally fast response (in the cyclotron and Larmor frequency range). 
The signature of Wilczek's anyons is a slight shift of the resistivity at the plateau of the IQHE.
Thus, the quest for high-$\mu_r$ materials at high frequencies, which is underway in the field of metamaterials, and the quest for anyons, are here found to be on the same avenue.  
\end{abstract}

\pacs{73.43.-f, 05.30.Pr, 03.65.Vf}
\maketitle

In 1982 Wilczek pointed out that a composite object consisting of a charged particle and a flux tube, referred to as anyon, would obey fractional statistics~\cite{Wilczek}. 
Anyons exist in a two-dimensional (2+1)D space~\cite{Wilczek, Leinaas}. 
They are Abelian or non-Abelian, depending on how their wavefunction evolves under particle exchange~\cite{Nayak}.
When two Abelian anyons are exchanged, the wavefunction acquires a phase factor.
Non-Abelian anyons can exist when the system has some degeneracy, 
such that exchange of two anyons corresponds to a unitary transformation 
of the wavefunction in the space of degenerate states~\cite{Nayak}. 
Apart from the fundamental interest in anyons, non-Abelian anyonic quasiparticles, if experimentally realized, could become the building blocks of fault-tolerant topological quantum computers~\cite{Nayak, Kitaev}. In this Letter, we propose an experimental realization of the original Wilczek's model for (Abelian) anyons~\cite{Wilczek}. 

\begin{figure}
	\mbox{\includegraphics[width=0.45\textwidth]{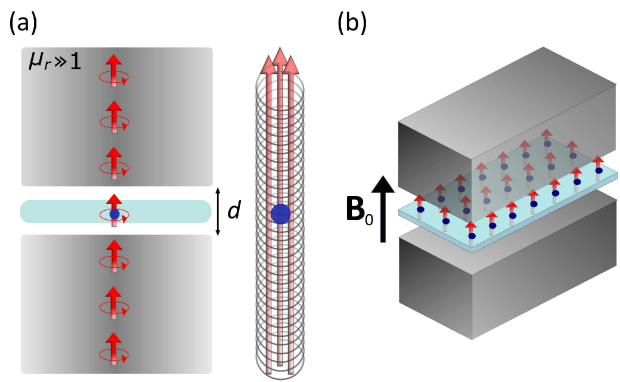}}	
	\caption{(Color online) The scheme which gives rise to Wilczek's flux-tube-charge composites. (a) A charged object with an intrinsic magnetic dipole (blue circle with a red arrow) induces an array of image magnetic dipole moments within high-$\mu_r$ blocks (shaded gray), which can be interpreted as a flux-tube-charge composite (central illustration). (b) A 2DEG in a uniform magnetic field ${\bf B}_0$ (in the IQHE state) is sandwiched between two blocks of high-$\mu_r$ material. Dipole magnetic moments of the electrons are aligned with ${\bf B}_0$, and behave as Wilczek's flux-tube-charge composites via the mechanism depicted in (a). 
	}
	\label{fig:scheme}
\end{figure}

Our scheme for creating charged flux tubes involves two semi-infinite blocks of a high permeability (high-$\mu_r$) material ($\mu_r{\gg} 1$), which are separated by some distance $d$, and a charged object with an intrinsic magnetic dipole moment. The object is located in the center of the slab between the high-$\mu_r$ materials, and its magnetic dipole moment is perpendicular to the surface of the blocks. The image potential of one such magnetic moment, arising from the high-$\mu_r$ material, creates an effective flux tube, thereby realizing flux-tube-charge composite, as illustrated in Fig.~\ref{fig:scheme}(a). The object could, for example, relate to an electron or a trapped ion, which have intrinsic magnetic moments. 

We use this scheme in a particular system to develop a proposal for a realization of Wilczek's anyons. 
Consider a two-dimensional electron gas (2DEG) placed in a perpendicular uniform magnetic field, which gives rise to the integer quantum Hall effect (IQHE)~\cite{Klitzing1980, Laughlin1981}. Suppose that we sandwich the 2DEG between two semi-infinite blocks of high-$\mu_r$ material, assumed to have a fast temporal response, see Fig.~\ref{fig:scheme}(b).
The electron spins (i.e., magnetic dipole moments) will be aligned due to the Zeeman effect, while the high-$\mu_r$ material will induce a flux tube attached to each electron.
For this system, we exploit the exact many-body wavefunction and calculate the Hall conductance. 
A signature of the presence of anyons in this system is striking: the Hall resistance at the plateau of the IQHE, which serves as a standard of electrical resistance~\cite{Klitzing1980,Jeckelmann2001,nist}, would be slightly shifted. 
We discuss possible implementations of the proposed system, the obstacles, and possible ways to overcome them.

In the search for the physical realization of anyons, quasiparticle excitations in two-dimensional interacting many-body systems play a major role~\cite{Nayak}. 
A paradigm of quasiparticles with fractional statistics are excitations in the fractional quantum Hall effect (FQHE)~\cite{Tsui, Laughlin1, Laughlin2, Halperin, Arovas, Camino}. 
The manifestation of both the IQHE and FQHE is a plateau in the Hall conductivity at $\nu e^2/h$, where the filling factor $\nu$ is an integer for the IQHE, and a fractional value for the FQHE. 
The key ingredients in the FQHE, described by the famous Laughlin state~\cite{Laughlin1, Laughlin2}, are 2D electrons in a strong uniform magnetic field~\cite{Tsui} and Coulomb interactions~\cite{Laughlin1, Laughlin2}.
In contrast, Coulomb interactions are not needed to explain the IQHE~\cite{Klitzing1980, Laughlin1981}; hence, we neglect them in our system. 
One way to explain the FQHE is via composite fermions~\cite{Jain1989, Jainbook, Lopez1991, Halperin1993}, where an electron is bound to an even number of the flux quanta, and the fractional Hall conductivity is interpreted as a manifestation of the IQHE of such composite fermions.
In the context of the QHE, anyons in a uniform magnetic field and corresponding wavefunctions have been studied~\cite{Johnson1990, Greiter1990, Grundberg1991, Dunne, Ouvry}. 
A slight shift of the Hall resistance at the IQHE plateau discussed here, which occurs from the conversion of electrons into Wilczek's flux-tube-charge composites - anyons - can be thought of as a variant of the composite fermions, however, with a completely different physical background.

\begin{figure}
	\mbox{\includegraphics[width=0.40\textwidth]{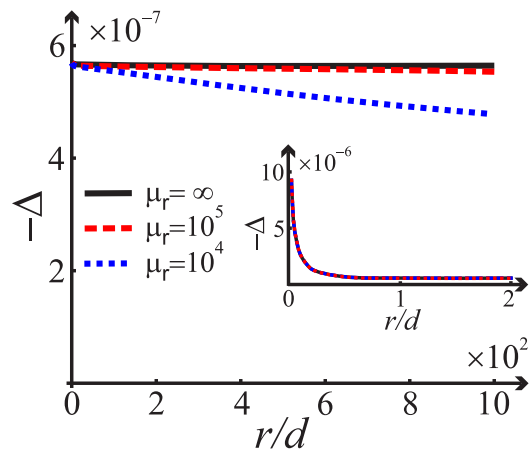}}
	\caption{(Color online) Parameter $-\Delta{=} -e \pi^{-1} \hbar^{-1} \oint {\bf A}\cdot {\bf dl}$ as a function $r/d$, for three values of $\mu_r$; $d{=}10$~nm (see text for details).
	}
	\label{fig:angle}
\end{figure}

Here we propose to convert electrons into anyons by introducing an electron-electron (e-e) vector potential mediated by the high-$\mu_r$ material. 
Our starting point is a 2DEG (in the $z{=}0$ plane) in a magnetic field ${\bf B}_0{=}B_0\hat{z}$ exhibiting IQHE. 
We assume that the electrons populate only the lowest Landau level, i.e., the filling factor is $\nu{=}1$. 
The two semi-infinite blocks of high-$\mu_r$ material with $\mu_r {\gg} 1$ are then introduced in the region $|z|{>}d/2$, see Fig. \ref{fig:scheme}(b).
The method of current images from classical electrodynamics models the influence of high-$\mu_r$ blocks on 
electrons, and allows one to calculate the magnetic vector potential ${\bf A}({\bf r})$ in 
the $|z|{<}d/2$ slab, due to magnetic dipole moment of a single electron~\cite{Jackson}. 
For a stationary magnetic dipole ${\bf m}{=}m\hat z$ located at the origin, in the limit 
$\mu_r{\rightarrow}\infty$, ${\bf A}({\bf r})$ is identical to that of an infinite array of magnetic 
moments deep within semi-infinite blocks. 
These virtual images are equal in magnitude and direction to the original magnetic moment, 
and equally spaced by $d$, as illustrated in Fig. \ref{fig:scheme}(a). 
Thus, for $r{=}|{\bf r}|$ sufficiently larger than $d$, an array of magnetic moments can be viewed as a flux tube with 
${\bf A}({\bf r}){\approx} \Phi / 2\pi r {\hat \phi}$, where the flux is $\Phi{=}\mu_0 m/d$. 
For a finite value of $\mu_r$, the vector potential in the $z{=}0$ plane is given by 
\begin{equation}
{\bf A}({\bf r})=\frac{\Phi r d}{4\pi}
\sum_{n\in \mathbb{Z}} \left(\frac{\mu_r-1}{\mu_r+1}\right)^{|n|} 
\frac{1}{(r^2+n^2d^2)^\frac{3}{2}}
{\hat \phi}.
\label{eq:vecseries}
\end{equation}
In order to estimate the validity of the approximation ${\bf A}({\bf r}){\approx} \Phi / 2\pi r {\hat \phi}$, in Fig.~\ref{fig:angle} we plot 
$\Delta{=}\frac{e}{\pi \hbar}\oint {\bf A}\cdot {\bf dl}$ as a function of $r$ and $\mu_r$ ($e<0$); 
the integral is taken around the circle of radius $r$ centered at the origin. 
Evidently, for $\mu_r{=}\infty$, $\Delta$ is essentially a constant independent of 
$r$ (except for $r{<}d$), verifying that the flux $\Phi{=}\oint {\bf A}\cdot {\bf dl}$ 
is concentrated close to the origin, and the approximation is excellent. 
For finite values of $\mu_r{=}10^4{-}10^5$, $\Delta$ changes very slowly over a large span of values of $r$ from $d$ up to the mean free path $l_{m.f.p.}$ in standard QHE samples~\cite{Beenakker}, which underpins the 
approximation in realistic circumstances. 
For concreteness, we plot Fig.~\ref{fig:angle} for $d{=}10$~nm, and $-\Delta$ is plotted up to $10000$~nm, 
but similar results are obtained for a span of values $d{=}10{-}100$~nm. 
We assume that the medium has sufficiently fast response, so that this picture is 
valid for a moving electron as well. 
This gives rise to the vector potential interactions between the electrons. 
The viability of the proposal and approximations are discussed below. 

If an electron encircles a fixed solenoid of flux $\Phi$, its wavefunction accumulates 
the Aharonov-Bohm phase $\text{exp}(ie\Phi/\hbar)$, but the same phase arises also from a quantum-mechanical solenoid orbiting around a fixed charge. 
Thus, the e-e vector potential mediated by the high-$\mu_r$ material 
is equivalent to that of a charge interacting with twice the flux in one flux tube~\cite{Wu, Goldhaber}, that is, the interaction is $2 e {\bf A}({\bf r}_i-{\bf r}_j)$, where
\begin{equation}
{\bf A}({\bf r}_i-{\bf r}_j)=\frac{\Phi}{2\pi}\frac{\hat{z}\times ({\bf r}_i-{\bf r}_j)}{|{\bf r}_i-{\bf r}_j|^2}.
\label{eq:potential}
\end{equation}
In the presence of a magnetic field, the electron energies of the up and down 
spins split due to the Zeeman effect. 
In large magnetic fields, large energies are needed to flip the spin; restricting to 
low energies, we can neglect electrons with magnetic moments opposite to that of the magnetic field. 
Under the assumptions and approximations stated above, 
the system is described by the Hamiltonian
\begin{equation}
H=\sum_{i=1}^n\frac{1}{2m}
\big[{\bf p} - e {\bf A}_0 ({\bf r}_i) 
- 2 e \sum_{j\neq i} {\bf A}({\bf r}_i-{\bf r}_j)\big]^2,
\label{eq:hamiltonian}
\end{equation}
with the symmetric gauge vector potential ${\bf A}_0=\frac{1}{2}{\bf B}_0\times{\bf r}$ for the constant magnetic field.

Electron-electron vector potential in (\ref{eq:hamiltonian}) is eliminated by a singular gauge transformation, 
\begin{equation}
\psi'({\bf r}_1,...,{\bf r}_n)=\prod_{i<j}
e^{-i \phi_{ij} \Delta}
\psi({\bf r}_1,...,{\bf r}_n),
\end{equation}
where $\phi_{ij}$ is the azimuthal angle of the relative vector ${\bf r}_i{-}{\bf r}_j$ and $\psi(\{{\bf r}_i\})$ is the fermionic wavefunction in the regular gauge. 
Note that the wavefunction $\psi'$ is multivalued. 
The wavefunction describing the electrons in the lowest Landau level, with the e-e vector potential, 
in the singular gauge is \cite{Johnson1990,Grundberg1991,Dunne,Ouvry}
\begin{equation}
\psi'(\{z_i\}\{z_i^*\})=\prod_{i<j}(z_i-z_j)^{\alpha}\text{exp}(-\frac{1}{4l_B^2}\sum_l|z_l|^2),
\label{eq:wavefunction}
\end{equation}
where we have introduced complex coordinates $z_i{=}x_i{+}iy_i$, the magnetic length $l_B{=}\sqrt{\hbar/eB_0}$, and the statistical parameter $\alpha{=}1 {-} \Delta$.
The energy of this state is independent of $\alpha$, $E{=}n\hbar\omega_c/2$, where $\omega_c{=}eB/m$ is the cyclotron frequency.

To calculate the Hall conductance in this system, we use the Laughlin's pumping argument in the Corbino ring geometry~\cite{Laughlin1981,Tong}. Suppose that we introduce an infinitely thin solenoid at $z{=}0$, and adiabatically increase the flux from $0$ up to $\Phi_0=2\pi\hbar/e$ (one flux quantum). 
The state (\ref{eq:wavefunction}) adiabatically evolves into 
\begin{equation}
\psi'_{0}(\{z_i\}\{z_i^*\})=\prod_i z_i \, \psi'(\{z_i\}\{z_i^*\}),
\label{eq:quasihole}
\end{equation}
which is an eigenstate of the system with the same energy. In this process, charge $q^{*}$ is pumped from the solenoid (at $z=0$) to the edge of the ring. It can be calculated from the single particle densities, $\rho_{\alpha}$ for the state (\ref{eq:wavefunction}), and  
$\rho_{\alpha,0}$ for the state (\ref{eq:quasihole}). 
The calculation is performed analytically in the thermodynamic limit $N{\rightarrow} \infty$ by using the plasma analogy, first introduced by Laughlin~\cite{Laughlin2} (see Refs.~\cite{Cappelli, Pasquier} for details),
\begin{equation}
\rho_{\alpha}(x,y)=\frac{1}{2\pi\alpha l_B^2}
\end{equation}
and 
\begin{equation}
\rho_{\alpha,0}(x,y)=\frac{1}{\alpha}\bigg[\frac{1}{2\pi l_B^2}-
\delta(x)\delta(y)\bigg].
\end{equation}
Evidently, the missing charge at $z=0$ is $q^{*}{=}e/\alpha$, which yields 
\begin{equation}
\sigma_{H} = q^{*}\frac{e}{h}
= \frac{1}{\alpha} \frac{e^2}{h}
\label{eq:Hall_conductivity}
\end{equation}
for the Hall conductivity. One can say that by attaching a flux $\Phi$ to every electron, one slightly reduces the Pauli repulsion between the electrons, which depletes the charge pumped to the edges by a factor $\alpha^{-1}$.

Thus, before we place the two high-$\mu_r$ blocks in the system, 
the initial value of the Hall conductivity is $\nu e^2/h$ with $\nu{=}1$ by assumption.
After placing the blocks, which induce the e-e vector potential, 
the Hall conductivity at the plateau shifts from $\nu{=}1$ to $1{/}\alpha{=}1{/}(1{-}\Delta){\approx} 1{+}\Delta$.
The shift $-\Delta$ is plotted in Fig.~\ref{fig:angle}, and it has the value $\sim 10^{-7}-10^{-6}$ (see Fig.~\ref{fig:angle}). 
Despite the fact that the shift is small, $\Delta \sigma_{H} {\sim} 10^{-7}\times e^2/h$, measurements indicate that the value of the quantized Hall resistance can be reproduced within a relative uncertainty of one part in $10^{10}$~\cite{nist}, meaning that the shift in the Hall conductance could be detectable as the signature of Wilczek's anyons. In addition, we note that as the e-e vector potential is introduced (a flux tube with flux $\Phi$ is adiabatically attached to every electron), according to the adiabatic principle developed by Greiter and Wilczek~\cite{Greiter1990}, the system remains gapped, i.e., incompressible quantum Hall states remain incompressible.

Now we discuss possible implementations of this system, the obstacles and possible routes to overcome them. We have assumed that the e-e vector potential picture is valid also for electrons moving in the 2DEG, even though it was derived for static electrons. In the classical picture, electrons exhibiting the Hall effect move in circular orbits with the cyclotron frequency, giving rise to oscillating fields that material should respond to. In the quantum picture, electrons are in the Landau level states. Recent experiments~\cite{Schattschneider2014} have demonstrated that the currents corresponding to electrons promoted in Landau level states oscillate at cyclotron ($\omega_c=eB/m^*$) and Larmor frequencies ($\Omega=eB/2m^*$), depending on the particular state; here $m^*$ is the effective mass of electrons. Therefore, we conclude that the demanded high-$\mu_r$ material should have a strong magnetic response in the frequency range corresponding to cyclotron motion. 
A typical system for the QHE is the interface of a GaAs/AlGaAs heterojunction where $m^*=0.067 m_e$~\cite{Lo}, and the frequencies are in the THz range.  Unfortunately, the magnetic response of most conventional materials is beginning to tail off in the GHz region~\cite{Pendry}. 
A few natural magnetic materials that respond above microwave frequencies have been reported, but the magnetic effects in these materials are typically weak (see Ref.~\cite{Yen} and references therein).
These restrictions can in principle be overcome by using metamaterials, artificial structures which can be constructed to have a strong effective magnetic response $\mu_{eff}(\omega)$ at high frequencies (GHz-THz)~\cite{Yen, Pendry, Merlin}.
Another advantage of using metamaterials in this context is that their response is usually not broadband. Therefore, a high-$\mu_r$ metamaterial at THz is likely to have low response (or none) at zero frequency (for example, see~\cite{Liberal2017}), and would not be affected by the constant magnetic field used to create IQHE state. 
One possible route for constructing a desirable metamaterial could be photonic doping, recently used to construct a material with effective $\mu_{eff}{\rightarrow} \infty$~\cite{Liberal2017} for polarization where magnetic field is parallel to the surface (here we demand that the magnetic field is perpendicular to the surface). 
The characteristic scale of the building constituents of the metamaterial should be smaller than the magnetic length $l_B$, so that the concept of the effective macroscopic permeability remains valid.
Another possibility to overcome the obstacle of fast material response is to reduce the Fermi velocity and thereby the cyclotron frequency by involving heavy fermion materials, in which electrons have a large enough effective mass.
The cyclotron frequency scales as $1{/}m^*$; thus, to bring the cyclotron frequency down to GHz range, by using typical numbers from above, the effective mass of the electrons should be $m^*{\sim} 10^2 m_e$.

An important parameter, which should be tuned to get the desired effect is the distance between the high-$\mu_r$ materials $d$. The flux tube approximation ${\bf A}({\bf r}){\approx} \Phi / 2\pi r {\hat \phi}$ for the vector potential of an electron, which is illustrated in Fig.~\ref{fig:angle}, is excellent already for $r{>}d$. 
We find that for values of $\mu_r{\sim} 10^4$ and larger, it is excellent up to $r{\sim}l_{m.f.p.}$ and more (this depends on $\mu_r$). It gives rise to the e-e interactions (\ref{eq:potential}). Hence, the average separation between electrons should be greater than $d$ for Eq. (\ref{eq:potential}) to apply. In standard IQHE experiments, the electron density is $10^{11}-10^{12}$~cm$^{-2}$, so that the average separation is of the order of $20$~nm, but in principle it could be larger.  
For larger values of $d$ (say $d{\sim}30{-}60$~nm), the flux tube approximation is even better at scales from $d$ to $l_{m.f.p.}$. However, the shift in the Hall conductivity $\Delta \sigma_H$, which is the signature of the effect, scales as $1/d$. Thus, we must find an appropriate value for $d$ smaller than the average separation between electrons, and small enough for the effect to be measurable, but large enough to be possible to sandwich a thin material with IQHE between two blocks of high-$\mu_r$ material. This is a viable task according to the parameters used in Fig.~\ref{fig:angle}. 
Moreover, assuming one could tune $d$ in an experiment, measurement yielding $\Delta \sigma_H \sim 1/d$ would be a clear evidence of Wilczek's anyons in this system.
Since the area of the IQHE sample is finite and $\nabla\cdot {\bf B}=0$, when $r {\rightarrow} \infty$, $-\Delta {\rightarrow} 0$. Thus, the high-$\mu_r$ materials should have a large aspect ratio (height much larger than the square root of the area), to properly steer the magnetic streamlines.

Before closing, we note that a promising possibility to observe Wilczek's flux tubes is to engineer 2D materials~\cite{Butler2013}. 
To this end, we propose to intercalate a metallic monolayer between two layers of hexagonal boron nitride (h-BN); this could be Li, K, Na or some other metallic monolayer~\cite{Doll1989, Sumiyoshi2010}.
The density-functional theory calculations for h-BN-Li-h-BN monolayer show structural stability and a parabolic band dispersion~\cite{Vito2018}.  
The principle of intercalation is here very similar to such intercalation in graphite, which has been extensively studied~\cite{Dresselhaus}. 
The h-BN - metallic monolayer - h-BN structure can in principle be sandwiched between two blocks of the high-$\mu_r$ material, thus constituting a candidate for observing anyons according to our scheme. 
Another route could be to grow a metallic monolayer on a film of a semiconductor as in Ref.~\cite{Eknapakul}, and place it between the high-$\mu_r$ blocks (the semiconductor should be sufficiently pure, not to conduct). 
Viable paths could also be conceived with layered dichalcogenides~\cite{Butler2013}.

For concreteness, our theoretical analysis above is based on the QHE with electrons in a 2D parabolic band. 
The most famous 2D material - graphene - has the conical band structure~\cite{Novoselov2005, Zhang2005, Castro-Neto}. 
However, graphene sandwiched between two blocks of high-$\mu_r$ material could also be a candidate for exploring (Wilczek-Dirac type) anyons according to the present proposal. Although the quantum Hall effect in graphene is distinctive, as it occurs at half-integer filling factors~\cite{Novoselov2005, Zhang2005}, the Landau-level wavefunctions for low-energy electrons in graphene have the same mathematical structure as in the 2DEG (up to the coefficients that enter these wavefunctions~\cite{Castro-Neto}).
Thus, we conjecture that the signature of Wilczek's flux tubes in this system would also be a small shift of the resistance at the plateau. Graphene also has the possibility to be strained~\cite{Si2016} and induce effective gauge fields, which is additional useful degree of freedom when tinkering with this system.

In conclusion, we have proposed a scheme for creating flux-tube-charge composites, which employs a material with high magnetic permeability $\mu_r$.  Thus, advances in developing high-$\mu_r$ metamaterials could lead to novel ways for creating anyons. 
We have calculated the Hall conductivity for a 2DEG in the IQHE regime, sandwiched between two semi-infinite blocks of high-$\mu_r$ metamaterial with a fast temporal response, and found that the Hall resistance at the plateau would exhibit a small but detectable shift, which is to some extent a striking consequence because it serves as a standard of electrical resistance~\cite{Klitzing1980, Jeckelmann2001, nist}. 
Finally, we would like to note that the quest for anyons is of broad interest and underway in many systems including ultracold atomic gases~\cite{Paredes,Burrello2010,Dai}, photonic lattices~\cite{Longhi2012} and quantum spin liquids~\cite{Klanjsek}. Our scheme for creating charged flux tubes has potential to be used in other systems such as trapped ions. 
Here we have addressed Abelian anyons. We believe that further studies inspired by this proposal could yield schemes for realizing non-Abelian anyons for topological quantum computing~\cite{Nayak}.

We acknowledge useful discussions with J. Jain, N. Lindner, M. Hafezi, T. Dub\v{c}ek, E. Tafra, M. Basleti\'{c}, M. Kralj, M. Po\v{z}ek, and V. Despoja.
This work was supported by the Croatian Science Foundation grant IP-2016-06-5885 SynthMagIA, and in part by the by the QuantiXLie Centre of Excellence, a project
co-financed by the Croatian Government and European Union through the
European Regional Development Fund - the Competitiveness and Cohesion
Operational Programme (Grant KK.01.1.1.01.0004). 
M.S. acknowledges support from the Army Research Office through the Institute for Soldier
Nanotechnologies under contract no. W911NF-13-D-0001.

\end{document}